# Design and Implementation Considerations for a Virtual File System Using an Inode Data Structure

Qin Sun
*Department of Computer Science*
*The Ohio State University*
*Columbus, OH, United States*
*sun.3301@osu.edu*

Grace McKenzie
*Department of Computer Science*
*The Ohio State University*
*Columbus, OH, United States*
mckenzie.401@osu.edu

Guanqun Song
Department of Computer Science
The Ohio State University
Columbus, OH, United States
song.2107@osu.edu

Ting Zhu
Department of Computer Science
The Ohio State University
Columbus, OH, United States
zhu.3445@osu.edu

*Abstract*—**Virtual file systems are a tool to centralize and mobilize a file system that could otherwise be complex and consist of multiple hierarchies, hard disks, and more. In this paper, we discuss the design of Unix-based file systems and how this type of file system layout using inode data structures and a disk emulator can be implemented as a single-file virtual file system in Linux. We explore the ways that virtual file systems are vulnerable to security attacks and introduce straightforward solutions that can be implemented to help prevent or mitigate the consequences of such attacks.**

*Keywords—file system, virtual disk, inode*

## I. INTRODUCTION

Virtual file systems are a useful way to abstract a concrete file system in order to simplify the view of the system into a unified entity which can also potentially be easily portable as well as customizable according to the needs of the user. A virtual file system acts as a layer on top of any operating system that provides a way to access data in a structured manner that avoids the complexities of the underlying system, which could be running multiple operating systems, or using multiple hard disks, for example. The virtual file system uses the same components for file management as a concrete file system, while at the same time allowing for abstraction to a single hierarchy rather than multiple hierarchies [4].

One basic type of a virtual file system is the single-file virtual file system, which can involve an emulator that encapsulates the functionality of a concrete file system and potentially also the disk layout, using software. This design is the basis for our project, in which we explore how the concrete file system can be modeled using inode data structures, as well as how to virtualize the disk and emulate its functionality using only a single file.

File systems are vulnerable to a wide variety of cyberattacks. Some of these attacks such as injection attacks, backdoor attacks, and unauthorized access to the system are preventable by security measures such as user authentication techniques, verifying user input, and file encryption [3].

Virtual file systems provide a unified and simplified view of complex file systems, allowing for easy portability and customization. They act as a layer over various operating systems, managing data in a structured way while abstracting the complexities of the underlying systems [4]. This abstraction is similar to how heterogeneous IoT devices manage diverse data types through concurrent communication [8-9], offering a unified interface over complex underlying structures. In addition, File systems, including virtual ones, face various cybersecurity threats. Solutions like user authentication, input verification, and file encryption [3] echo the security protocols used in IoT networks [10-11], ensuring integrity and preventing unauthorized access.

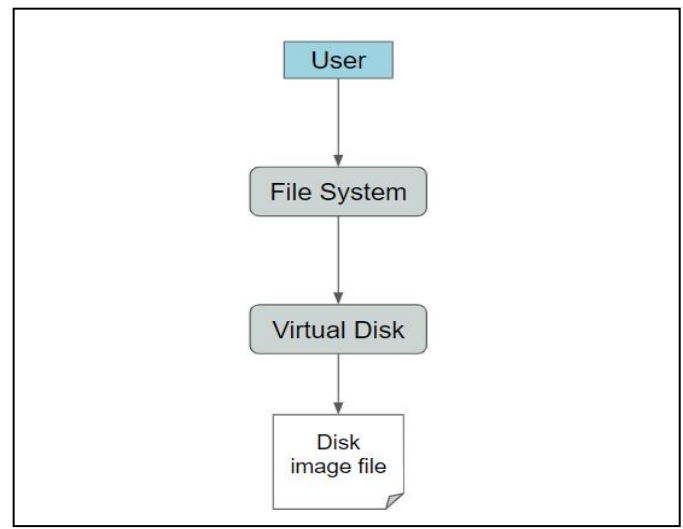

Fig. 1. Program flow chart showing the flow of input. (*figure caption*)

Virtual file systems provide an essential abstraction layer over various operating systems, streamlining data management in a way that resonates with the complexities of IoT environments. This abstraction is crucial in handling diverse data types and communication protocols, prevalent in heterogeneous IoT networks [8-11, 15-19]. Our exploration of inode data structures in virtual file systems draws parallels with efficient data dissemination and aggregation strategies in IoT networks [16, 17].

In addressing cybersecurity threats to virtual file systems, we take cues from secured protocols developed for IoT devices [22, 23]. The development of robust security measures, such as

```
struct super_block
{
    int no_of_blocks_used_by_superblock = ceil(((float)sizeof(super_block)) / BLOCK_SIZE);

    int no_of_blocks_used_by_file_inode_mapping = ceil(((float)sizeof(struct file_to_inode_mapping) * NO_OF_INODES) / BLOCK_SIZE);

    int starting_index_of_inodes = no_of_blocks_used_by_superblock + no_of_blocks_used_by_file_inode_mapping;

    int no_of_blocks_used_to_store_inodes = ceil(((float)(NO_OF_INODES * sizeof(struct inode))) / BLOCK_SIZE);

    int starting_index_of_data_blocks = no_of_blocks_used_by_superblock + no_of_blocks_used_by_file_inode_mapping +
        no_of_blocks_used_to_store_inodes;

    int total_no_of_available_blocks = DISK_BLOCKS - starting_index_of_data_blocks;

    bool inode_freelist[NO_OF_INODES];
    bool datablock_freelist[DISK_BLOCKS];
};
```

Fig2. Superblock structure definition. (figure caption)

user authentication and file encryption, reflects similar strategies in ensuring data integrity and preventing unauthorized access in IoT environments [24, 31, 41].

Our project explores the use of inode data structures in modeling concrete file systems, alongside virtualizing disk functionalities using a single file.

## II. DISCUSSION ON UNIX FILE SYSTEM DESIGN

### A. Inode Data Structure and Pointer Structures

An inode data structure comprises multiple inode pointers which point to data blocks in other locations within the disk. In most cases, direct pointers make up the majority of the inode pointers. Direct pointers are pointers which directly point to blocks that store the file's data. Although direct pointers are the major component in the inode pointers and are efficient to look up, one significant drawback of direct pointers is that it is hard to keep track of a large size of data using direct pointers in limited inode space. To solve this issue, indirect pointers are introduced into the inode data structure. In contrast to the fact that there are only a small number of indirect pointers in the inode pointer array, they store much more data than direct pointers do. Unlike direct pointers, indirect pointers don't point directly into blocks that store data. Instead, indirect pointers point to blocks that store pointers pointing to data blocks. With this structure, the number of data blocks that an indirect pointer could point to is exponentially increased, and the indirect pointer can be enhanced further by introducing more layers of indirect pointing. This hierarchy design greatly augments the capability of inode data structure to store large files [7].

## III. IMPLEMENTATION OF AN INODE-BASED VIRTUAL FILE SYSTEM IN LINUX

### A. Inode Data Structure Modeling

In our virtual file system, we implemented the inode structure as an integer value storing file size plus an integer array storing pointers. The pointer array is composed of 10 direct pointers, one single indirect pointer and one double indirect pointer. Each integer occupies four bytes of storage, so one inode structure takes up 52 bytes in total.

```
1 : create disk
2 : mount disk
9 : exit
2
Enter diskname :
test/mydisk
Disk is mounted!!!
=========================
1 : create file
2 : open file
3 : read file
4 : write file
5 : append file
6 : close file
7 : delete file
8 : list of files
9 : list of opened files
10: unmount
=========================
```

Fig 3. Superblock structure definition. (*figure caption*)

### B. Disk Emulation

In our virtual file system, we use a single file to emulate the behavior of a disk in the Linux system. When a user requests to create a disk, the file system will first check if a disk with the same name already exists. If it does, the file system will return an error prompt. Otherwise, the file system will start the process of creating a new file and initializing its space. The file system initializes the file space with a buffer which size equals to the preset block size. The first step of creating a disk file is to set the space to null value block by block with the buffer. Then we create a temporary superblock structure for this specific file disk and initialize the attributes within the super block structure.

After that, we copy crucial structures such as super block, file inode mapping and inodes into the file disk in the appropriate sequence as designed in super block structure.

In order to use the created disk, the user must then choose the option to mount it. The user is prompted to enter the name of the disk they wish to mount, then is notified of a successful or unsuccessful completion of the mount action. If successful, the user is presented with a list of file operations to perform with associated numbers, as well as the option to unmount the disk (See Fig. 3 for a demonstration). After unmounting the disk, the user may return to the previous set of options which includes creating another disk, mounting a disk, or exiting the program.

*C. Basic File Operations*

Once the user has successfully mounted their disk, they have the option to choose from several actions to perform. We have defined the following basic file operations in our implementation of a virtual file system so that users can expect to be able to manipulate their files with nearly the same basic functionality as the Linux file system.

In order to have any form of access to a particular file, the user must select the option to open a file and provide the filename of the file they wish to open. Once the user provides this information, they must then select an option to add a mode to the file: read, write, or append. As is the same in existing established file systems, our implementation defines reading, writing, and appending to a file as viewing a file, inserting text into a file, and adding text to the end of a file, respectively. Once the user makes the mode selection, a file descriptor is assigned to the file to designate that it has been successfully opened. The file descriptor is printed to the program output because the user must know this descriptor and use it as input to perform additional operations on this file.

Fig 4. The user's command line view after choosing the option to mount a disk. (*figure caption*)

Fig 5. The user's command line view after choosing to open a file, then choosing to write to it by entering its file descriptor.

To close a file, the user gives as input the file descriptor that was assigned to the file when it was opened. The file descriptor is then cleared of its mapping to the given file and the user is notified that the file has been closed successfully.

If the user wants to create a file, they enter the appropriate option and will be prompted to enter a filename. The filename is stored and a success message is printed provided the given filename is valid. The file system looks for the next inode that is marked as free and a free data block is assigned to it. The free inode and data block are changed to indicate they are currently in use. The file to inode mappings are updated accordingly.

Deleting a file is the reverse of the process to create a file. The mappings for the file to the inode are removed, the data block in use is marked as free, and the inode mapped to the file is marked as free, as well.

The user can choose to read, write, or append to an opened file. Upon making the desired selection, the user is prompted to enter a file descriptor which is only valid for open files. To read a file, the file system checks for all data blocks that contain the data for a given file. The data is returned to the user.

To write or append to a file, the file system calculates the current position of the correct block to write to for the given file. The user inputs the text they wish to write or append to their file, followed by the special string EOF that indicates the user has completed their text entry (See Fig. 4 for a demonstration). Provided that the size of the text input by the user is small enough to fit into the remaining data blocks of the system, the text is mapped to the appropriate data block.

If the user would like to see all files they have stored in their mounted disk, they can choose the option to list files and

```
List of All files
bar.c with inode : 1
foo.c with inode : 0
main.java with inode : 2
program.cpp with inode : 3
=========================
1 : create file
2 : open file
3 : read file
4 : write file
5 : append file
6 : close file
7 : delete file
8 : list of files
9 : list of opened files
10: unmount
=========================
```

Fig 6. The user's command line view after choosing the option to list all files stored in the disk. (figure caption)

the program will return the filenames to output (see Fig. 5 for a demonstration). The user may also want to see only the files that are currently open. To do so, they can select the list opened files option, as well.

## IV. SURVEY OF SECURITY IN FILE SYSTEMS

### A. Security Threats for File Systems

Each type of file system has its set of drawbacks when it comes to protecting against security threats. The protection of user data is a primary objective for file systems; however, file systems remain vulnerable to certain kinds of attacks including those utilizing malware, and those that exploit bugs. Such attacks can produce drastic consequences for not only the user data stored in the file system but also for the system itself.

Bugs in the software for a virtual file system can allow a malicious user to perform actions that are not intended to be part of the program. For example, a backdoor attack can give an attacker read or write access to a user's files that the attacker would otherwise not have had permission to view. In addition, an injection attack can allow the attacker to break the program if they have inserted commands that perform illegal actions, such as deleting important files that they do not have permission to delete. Attackers can also exploit the file systems' memory access. Critical errors related to memory management such as null pointer exceptions and buffer overflows can be forced by attackers looking for ways to corrupt memory or cause a crash. Some file systems rely on network access to shared data, such as distributed file systems. Multiple users and multiple servers accessing shared data creates additional security vulnerabilities for the file system [3].

### B. Potential Solutions for Security Concerns

There are several proposed solutions to some of the aforementioned security threats to file systems. One of these is user authentication. A basic concrete file system allows a user who has logged into their profile on the operating system to perform disk and file operations. In the event that an unauthorized user has gained access to another user's profile, they then have access to their files, as well. In this instance, it may be beneficial to have another layer of password protection for accessing the user's files. A virtual file system employing user authentication may be helpful, because it can require authentication to mount or delete a disk, for example.

When there are multiple users that need to access files in a file system, this introduces the need for access control lists, which associate permissions with a resource for any given user. This solution is particularly beneficial for distributed file systems and other file systems involving shared resources [1].

Encryption is another layer of security that can be applied to file systems to enhance access control and further protect sensitive file content and metadata. Attackers will have extreme difficulty decrypting disks and files, and access to the physical disk will not be useful to them [2].

There are two main categories of encryption for file systems: encryption at the file system level, and encryption at the disk level. File system-level encryption includes cryptographic file systems, which can be used to add protection to a primary file system, and involves the file system itself encrypting files or directories. This typically would not include the encryption of file metadata. Disk-level encryption, as opposed to file system-level encryption, is used to encrypt the entire contents of a disk, sometimes including the master boot record. A more secure file system can be achieved by using disk encryption and file system encryption together to utilize the benefits of both, and avoid the drawback of disk encryption in that it typically requires only one key to encrypt and decrypt the whole disk [5].

### C. Implementing Select Solutions

Our implementation of a virtual file system in Linux aims to provide some basic security features so that attacks such as those previously mentioned can be prevented. For protection against injection attacks similar to those mentioned above, we would implement a form of checking in user handling functions to ensure that only valid commands are input by the user, and any invalid or extraneous text will cause the input to be rejected and the action terminated. In order to guard against unauthorized users gaining access to a system and therefore gaining unauthorized access to the virtual file system, we would implement a login process prompting for a username and password in order to access the disk, or mount the disk. It would also then be necessary to have new actions for the user, to set their initial username and password and to replace a forgotten password. Additionally, it would be possible to implement encryption into our file system in such a way that the user can choose a command to encrypt or decrypt a given file using their password or another private string as key. These are simple adjustments to the file system that would help safeguard against several kinds of attacks thus improving the functionality of the program so users are more likely to benefit from using it [1].

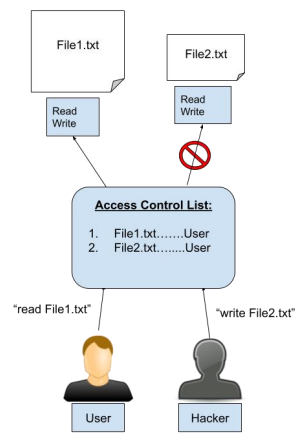

Fig 7. A visualization of file permissions and an access control list to add another layer of protection against unauthorized users in our implementation of a virtual file system. (*figure caption*)

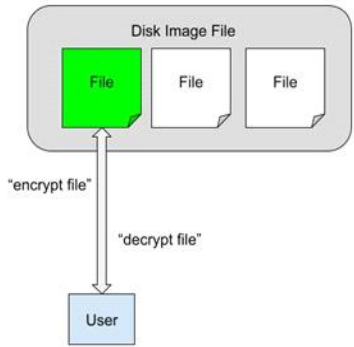

Fig 8. A visualization of file system-level encryption in our implementation of a virtual file system. (*figure caption*)

## V. CONCLUSION AND FUTURE WORK

Virtual file systems are a flexible tool for simplifying the way we view and access file systems, adapting them to demands such as shared resources, and making them more portable and easier to use. One type of virtual file system is a single-file virtual file system, which treats a single file as a disk image file. Users of single-file virtual file systems benefit from the portability and centralized structure of the file system design. Our implementation of a virtual file system uses this approach, along with the inode data structures and pointer system used by Unix-based file systems. The inode data structures and pointers are modeled using structures and arrays of pointers to inode structures. The virtual disk is emulated by dividing the given disk image file into blocks. To perform file operations, files are assigned free data blocks via direct pointers, indirect pointer, and double indirect pointer as necessary. Basic file operations implemented include create, delete, open, close, read, write, and append. Disk operations implemented include create, mount, and unmount. Users can make selections to perform commands on their disk and files by entering the associated index for each action into the command line.

Security features such as user authentication by login with username and password, user input verification, and file encryption and decryption using a user-defined password as a key were discussed as solutions to several possible security threats to virtual file systems, namely: unauthorized users gaining access to a system and thus gaining access to the virtual file system, and injection attacks and backdoor attacks due to software bugs.

The initial implementation of the inode-based virtual file system is intended to explore the modeling of inode data structures and how this would be implemented using a single disk file on top of Linux. The program would benefit from expanding to include implementation of additional file operations to enhance usability and practicality.

We intend to continue development of the virtual file system by completing implementation and testing of security features including user authentication in the form of login with username and password; encryption and decryption for files and related user input handling; and parsing user input for the purpose of rejecting invalid commands and other text entered by the user. Rigorous testing is necessary to discover and locate any bugs in the program that generate holes in the security of the file system, or other major impacts on functionality of the program, such as the integrity of the data.

Once we have completed sufficient implementation of security features to the virtual file system that provide adequate protection to be considered useful to an average user, we will continue to improve the user experience of the program. This will include providing detailed error messages to aid users in their understanding of what and why the action has caused an error. We will also increase the level of detail provided in any instruction messages that are printed out for the user, to help prevent potential confusion and mistakes. Due to the nature of using a single disk file to emulate the functionality of a physical disk, there may be significant

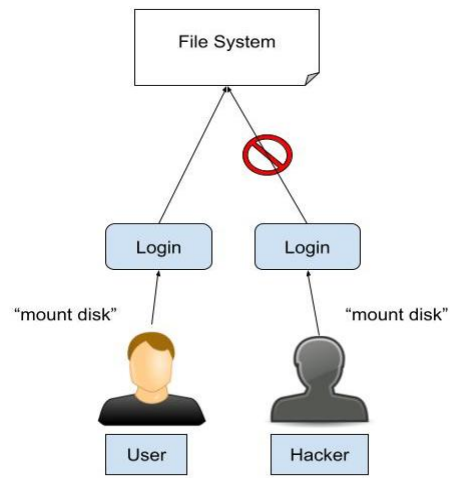

Fig 9. A visualization of user authentication via username and password login for mounting a disk in our implementation of a virtual file system. (*figure caption*)

running time delays when performing actions such as creating and mounting disks. Loops containing array accesses are a necessary part of initializing data structures for the virtual disk. We will explore ways to reduce the runtime of such procedures, as well as implement a loading or progress indicator to show the user that the process is being performed and tell the user the status of the process, including whether there are errors. Following this improvement is the need to disallow input while waiting for process completion, to avoid fatal errors for the program and confusion for the user.

## VI. EXTENDED CONSIDERATION

Our inode-based virtual file system initiative aligns with advanced data management and communication techniques in IoT [9-21]. Future enhancements in file operations and user interfaces will incorporate principles from heterogeneous computing systems [14] and leverage machine learning algorithms [25, 27, 30] to improve user interaction and system efficiency.

Expanding security features will involve integrating IoT security strategies [22, 23, 31] and adapting machine learning-based secure communication [22, 38]. This approach aims to bolster our system against evolving cybersecurity threats, drawing inspiration from the secure management of IoT networks and smart health systems [35, 36, 40].

Optimization of runtime processes will look towards efficient data handling techniques found in IoT and smart system networks [32-35, 39, 42, 43], enhancing user experience and system responsiveness. The goal is to achieve a level of efficiency and user-friendliness that mirrors advanced computational systems and smart applications [26, 28, 38].

In conclusion, our project will continuously evolve, integrating state-of-the-art concepts from IoT, smart systems, and machine learning. This integration ensures our virtual file system remains robust, user-friendly, and in line with contemporary technological advancements.